\documentclass[twocolumn,prd,footinbib,aps,floats,floatfix,amsmath,amssymb,longbibliography,
secnumarabic]{revtex4} %
\usepackage[utf8]{inputenc}
\usepackage{verbatim}
\usepackage{graphicx}
\usepackage{xcolor}
\usepackage{amsmath}
\usepackage{amssymb}
\usepackage[a4paper, total={6.5in, 10in}]{geometry}
\usepackage{hyperref}
\usepackage{mathtools}
\usepackage{import}
\usepackage{mathtools}
\usepackage{accents}

\usepackage{slashed}

\newcommand{\be}{\begin{equation}}
\newcommand{\ee}{\end{equation}}
\newcommand{\ba}{\begin{array}}
\newcommand{\ea}{\end{array}}
\newcommand{\bea}{\begin{eqnarray}}
\newcommand{\eea}{\end{eqnarray}}

\def\resp#1{{\color{black} #1}}

\usepackage[utf8]{inputenc}

\begin{document}
\title{Blazar constraints on neutrino-dark matter scattering}
\author{James M. Cline}
\affiliation{McGill University, Department of Physics, 3600 University Street,
Montr\'eal, QC H3A2T8 Canada}
\author{Shan Gao}
\affiliation{McGill University, Department of Physics, 3600 University Street,
Montr\'eal, QC H3A2T8 Canada}
\author{Fangyi Guo}
\affiliation{McGill University, Department of Physics, 3600 University Street,
Montr\'eal, QC H3A2T8 Canada}
\author{Zhongan Lin}
\affiliation{McGill University, Department of Physics, 3600 University Street,
Montr\'eal, QC H3A2T8 Canada}
\author{Shiyan Liu}
\affiliation{McGill University, Department of Physics, 3600 University Street,
Montr\'eal, QC H3A2T8 Canada}
\author{Matteo Puel}
\affiliation{McGill University, Department of Physics, 3600 University Street,
Montr\'eal, QC H3A2T8 Canada}
\author{Phillip Todd}
\affiliation{McGill University, Department of Physics, 3600 University Street,
Montr\'eal, QC H3A2T8 Canada}
\author{Tianzhuo Xiao}
\affiliation{McGill University, Department of Physics, 3600 University Street,
Montr\'eal, QC H3A2T8 Canada}

\begin{abstract}
Neutrino emission in coincidence with gamma rays has been observed
from the blazar TXS 0506+056 by the IceCube telescope.  Neutrinos from the blazar had to pass through a dense spike of dark matter (DM) surrounding the central black hole.  The observation of such a neutrino implies
new upper bounds on the neutrino-DM scattering cross section as a function of the DM mass.  The constraint is stronger than existing ones for a range of DM masses, if the cross section rises linearly with energy.
For constant cross sections, competitive bounds are also possible, depending on details of the DM spike.
\end{abstract}

\maketitle

{\bf 1. Introduction.} 
The possible interactions of dark matter (DM) with ordinary matter
have been constrained in many ways.  The most challenging category is
DM-neutrino interactions, due to the difficulty of observing neutrinos.  A promising strategy is to consider astrophysical sources of high-energy neutrinos, that could accelerate light DM particles
to energies that would make them detectable in ground-based DM and neutrino search
experiments \cite{Wang:2021jic,Ghosh:2021vkt,Granelli:2022ysi}.  This only works if, in addition to DM-$\nu$
interactions, there can also be scattering of DM from nuclei or
electrons in the detector.

A more model independent strategy is to use the fact that a  \resp{290
TeV} neutrino, \resp{known as event IC-170922A,} has been observed by
the IceCube experiment and was identified as coming from the blazar
TXS 0506+056~\cite{IceCube:2018dnn}.  Ref.\ \cite{Choi:2019ixb} set
limits on the DM-$\nu$ scattering cross section using the fact that
the neutrino had to pass through cosmological and galactic DM between
the blazar and the Earth. In this work, we derive stronger limits,
using the fact that the neutrino  also had to traverse the dense DM
spike surrounding the supermassive black hole powering TXS 0506+056.

IceCube \resp{additionally} reported a statistical excess of lower
energy neutrinos prior the 2017 flare of TXS
0506+056~\cite{IceCube:2018cha}, but the claimed excess is too large
to be explained by \resp{state-of-the-art one-zone} blazar
models, \resp{likely requiring more complicated
modelling}~\cite{Keivani:2018rnh,Murase:2018iyl,Reimer:2018vvw,Rodrigues:2018tku,Petropoulou:2019zqp,Gasparyan:2021oad}.
\resp{Hence} we do not include it in the present analysis. \resp{
There have also been several candidate associations between
neutrinos detected by IceCube and known $\gamma$-ray blazars
subsequent to
IC-170922A
(e.g.~\cite{Rodrigues:2020fbu,Giommi:2020viy,Fermi-LAT:2019hte,Kadler:2016ygj,Sahakyan:2022nbz,IceCube:2021slf,Giommi:2020hbx,Franckowiak:2020qrq}).
Since none of them have been confirmed by the IceCube collaboration,
we do not include them in this study.}

{\bf 2. Expected neutrino events.} 
We start by describing the theoretical models of neutrino emission
from blazars and the expected flux from TXS 0506+056.  The observed spectra of electromagnetic emission from blazars is well described by
lepto-hadronic models \cite{Mucke:2000rn,Cerruti:2018tmc,Gao:2018mnu,Petropoulou:2019zqp,Gasparyan:2021oad}, in which protons and electrons are shock-accelerated to create a relativistic jet, in a magnetized region that produces synchrotron radiation.   The jet extends to distances $\sim 10^{11}\,$km \cite{Cerruti:2018tmc,Gasparyan:2021oad},
around $1000$ times smaller than the extent of the DM spike to be
described in Section 3. 
Proton-photon interactions in the jet produce pions, whose decays are the source of high-energy neutrinos.
 
Purely hadronic models are also able to fit the combined
electromagnetic spectra at optical, X-ray and gamma-ray frequencies,
but they lead to \resp{either a detectable neutrino flux at  much
higher energies or a negligible low flux at energies compatible with}
IC-170922A
\cite{Cerruti:2018tmc,Gao:2018mnu}; hence we focus on lepto-hadronic models
\resp{in the following. The impact of different choices is discussed
at the end of section 4}. Under the steady state
approximation, the hadronic model of Ref.\ \cite{Cerruti:2018tmc} predicts a neutrino flux
between $E_\nu \sim 100\,$TeV and 10\,EeV, that peaks at a value
$E_\nu \sim 10\,$PeV, which is orders of magnitude higher than
IC-170922A.  We find that the probability of observing a neutrino with
energy $\lesssim 300\,$TeV is $\sim 3\,\%$ in this model.  Hence we
consider it to be disfavored for explaining IC-170922A.  

On the other hand, the neutrino flux predicted by  \resp{the
lepto-hadronic model of} Ref.\ \cite{Gasparyan:2021oad}, based on a
fully time-dependent approach, peaks near $E_\nu=100\,$TeV, and is
compatible with the observation. \resp{Within the quasi-two neutrino
oscillation approximation~\cite{Gasparyan:2021oad},} the flux is well-fit
by the formula

\be
   \log_{10}\Phi_{\nu}(E_\nu)  = -F_0 - {F_1\,x\over 1 + F_2|x|^{F_3}}  
   \label{soprano-flux}
\ee
with \resp{$F_0 = 13.22, F_1 = 1.498, F_2 = -0.00167, F_3 = 4.119$, and $x=\log_{10}(E_\nu/{\rm TeV}) \in[-1.2,\,4.2]$}.
The expected number of \resp{muon} neutrino events observed at IceCube is given by
\be
    N_{\rm pred} = t_{\rm obs} \int dE_\nu\, \Phi_\nu(E_\nu)\, A_{\rm eff}(E_\nu)\,,
    \label{Npred}
\ee
where $t_{\rm obs}$ is the time interval of observation, $\Phi_\nu$
is the predicted neutrino flux from the blazar, and $A_{\rm eff}$
is an effective area for detection, which depends on the geometry of the source direction and $E_\nu$, and encodes the probability for a neutrino to convert to a muon through weak interactions.  Data for
$A_{\rm eff}$ from TXS 0506+056 is provided by IceCube \cite{icecube-data}.\footnote{The effective area can be fit in the region $x\in [-1,\, 6]$ by  $\log_{10} {A_{\rm eff}/ {\rm cm}^2} \cong
        3.57 + 2.007\, x - 0.5263\, x^2
        + 0.0922\, x^3 - 0.0072 \, x^4$}\ \ 
For the campaign IC86c during JD (Julian day) $57161-58057$ that 
observed IC-170922A, $t_{\rm obs} = 898\,$d, 
and the reconstructed energy was $E_\nu = 290\,$TeV.
This yields \resp{$N_{\rm pred} \approx 2.0$} 
from the flux (\ref{soprano-flux}), compatible with the observed event.  We adopt this as the input model for
constraining the DM-$\nu$ cross section in the following.

{\bf 3.\ Dark matter spike.} 
The overdensity of DM surrounding the central black hole plays a
crucial role for contraining $\nu$-DM scattering from the blazar.  The possibility of adiabatic accretion of DM around the
black hole (BH) was first considered by  Gondolo and Silk in Ref.\
\cite{Gondolo:1999ef}. They derived an inner radius for the spike of
$r_i = 4 R_S$, where $R_S = 2 G M_{BH}$ is the BH Schwarzschild
radius, and an outer profile $\rho'(r) \cong N\,(1 - 4 R_S /r)^3\,
r^{-\alpha}$ with $\alpha = (9-2\gamma)/(4-\gamma) \in [2.25,\, 2.5]$,
depending on the inner cusp of the initial DM halo density, $\rho\sim
r^{-\gamma}$, with $0\le\gamma\le 2$.   The normalization $N$ of
$\rho'$ can be determined using the finding that the mass of the spike
is of the same order as $M_{BH}$ \cite{Ullio:2001fb}, $4\pi
\int_{r_i}^{r_o} dr\, r^2 \rho' \cong M_{BH}$, within a radius of
typical size $r_o \cong 10^5\, R_S$ \cite{Gorchtein:2010xa}. The BH
mass of the blazar TXS 0506+056 is estimated to be $3.09 \times
10^{8}\,\,M_{\odot}$~\cite{Padovani:2019xcv}.  In Ref.\
\cite{Gnedin:2003rj}, it was argued that gravitational scattering of
DM with stars in the central region would lead to dynamical relaxation
to a less cuspy profile with $\alpha = 3/2$; hence we also consider
this possibility below.

The spike
density is reduced relative to these initial profiles if there is subsequent DM annihilation, leading to a maximum density of $\rho_c = m_\chi/(\langle\sigma_a v\rangle \,t_{BH})$, where $m_\chi$ is the DM mass, $\langle\sigma_a v\rangle$ is an effective annihilation cross section, and $t_{BH}$ is the age of the BH.  The spike density then becomes $\rho_\chi = \rho_c\rho'/(\rho_c+\rho')$.  The quantity $\langle\sigma_a v\rangle$
is ``effective'' in the sense that it could be negligible even if the
actual annihilation cross section is large.  This would be the case for asymmetric dark matter, in which the symmetric component has completely annihilated away in the early universe.  Then annihilations would have no effect at later times, when the DM spike is formed.  To illustrate the range of possible outcomes from varying $\langle\sigma_a v\rangle$, we follow Ref. \cite{Granelli:2022ysi} by considering three benchmark models BM1-BM3, in which $\langle\sigma_a v\rangle = (0,\, 0.01,\, 3)\times 10^{-26}$\,cm$^3$/s, respectively,
and $t_{BH} = 10^9\,$yrs. These models assumed $\alpha=7/3$ in $\rho'\sim r^{-\alpha}$.  We
also consider models BM$1'$-BM$3'$ using the less cuspy value $\alpha = 3/2$.

The probability for neutrinos to scatter from DM in the spike
depends on the DM column density, 
\be
     \Sigma_{\chi} = 
     \int_{R_{\rm em}}^\infty
     dr\, \rho_\chi \cong A_\Sigma \left(m_\chi
        \over 1\,{\rm MeV}\right)^{1-B_\Sigma} \,\,{\rm MeV} 
        \label{Sigma_eq}\,,
\ee
\resp{where $R_{\rm em} \approx R'\,\delta \sim 2 \times 10^{17}$ cm is the distance
from the central BH to the position in the jet where neutrinos and photons are
likely to be produced~\cite{Padovani:2019xcv}. $R' \sim 10^{16}$ cm is the comoving size of the spherical emission region and $\delta \sim 20$ is the Doppler factor for the lepto-hadronic model of Ref.~\cite{Gasparyan:2021oad}.}
One finds that $\Sigma_{\chi}/m_\chi$ can be accurately fit by a power law,
$\Sigma_{\chi}/m_\chi = A_\Sigma\, ({\rm MeV}/m_\chi)^{B_\Sigma}$, with 
$B_\Sigma = 1$ for the case of $\langle\sigma_a v\rangle = 0$, and  a fractional
power when annihilation occurs.   The parameters  $A_\Sigma,\,B_\Sigma$
for the benchmark models are given in table \ref{tab1}. 
Although the DM spike does not
extend to arbitrary distances, the integral in (\ref{Sigma_eq}) converges around
$10\,R_S$ in the case of no DM annihilation,  and at larger radii  $\sim
(10^6-10^8)\, R_S$ for the cases with annihilation.

\begin{table}
\centering
\resizebox{8cm}{!}{\begin{tabular}{ |c|| c | c | c |c||c|c|c|c|} 
 \hline
 
$\langle\sigma_a v\rangle$ & Model & $\alpha$ &$\log_{10}{A_\Sigma}$ & $B_\Sigma$ & Model & $\alpha$ & $\log_{10}{A_\Sigma}$ & $B_\Sigma$ \\
 \hline
 $0$ & BM1 & $7/3$ & $31.4$ & 1 & BM$1'$ & $3/2$ & $31.9$ & 1 \\
 \hline
$0.01$ &  BM2  & $7/3$ & $30.0$ & $0.48$  & BM$2'$ & $3/2$ & $30.8$ & $0.73$ \\
 \hline
$3$ &  BM3  & $7/3$ & $28.7$ & $0.43$  & BM$3'$ & $3/2$ & $29.5$ & $0.66$ \\
 \hline
\end{tabular}}
\caption{Normalization $A_\Sigma$ and exponent $B_\Sigma$ of power law fit to DM spike column density per mass $\Sigma/m_\chi$; see Eq.\ (\ref{Sigma_eq}).  Models are distinguished by different values of the effective DM annihilation cross section (in units of $10^{-26}$\,cm$^3$/s) and the spike profile exponent $\alpha$.  $A_\Sigma$ is in units of cm$^{-2}$.}
\label{tab1}
\end{table}

\begin{figure}[t]
   \centerline{\includegraphics[scale=0.335]{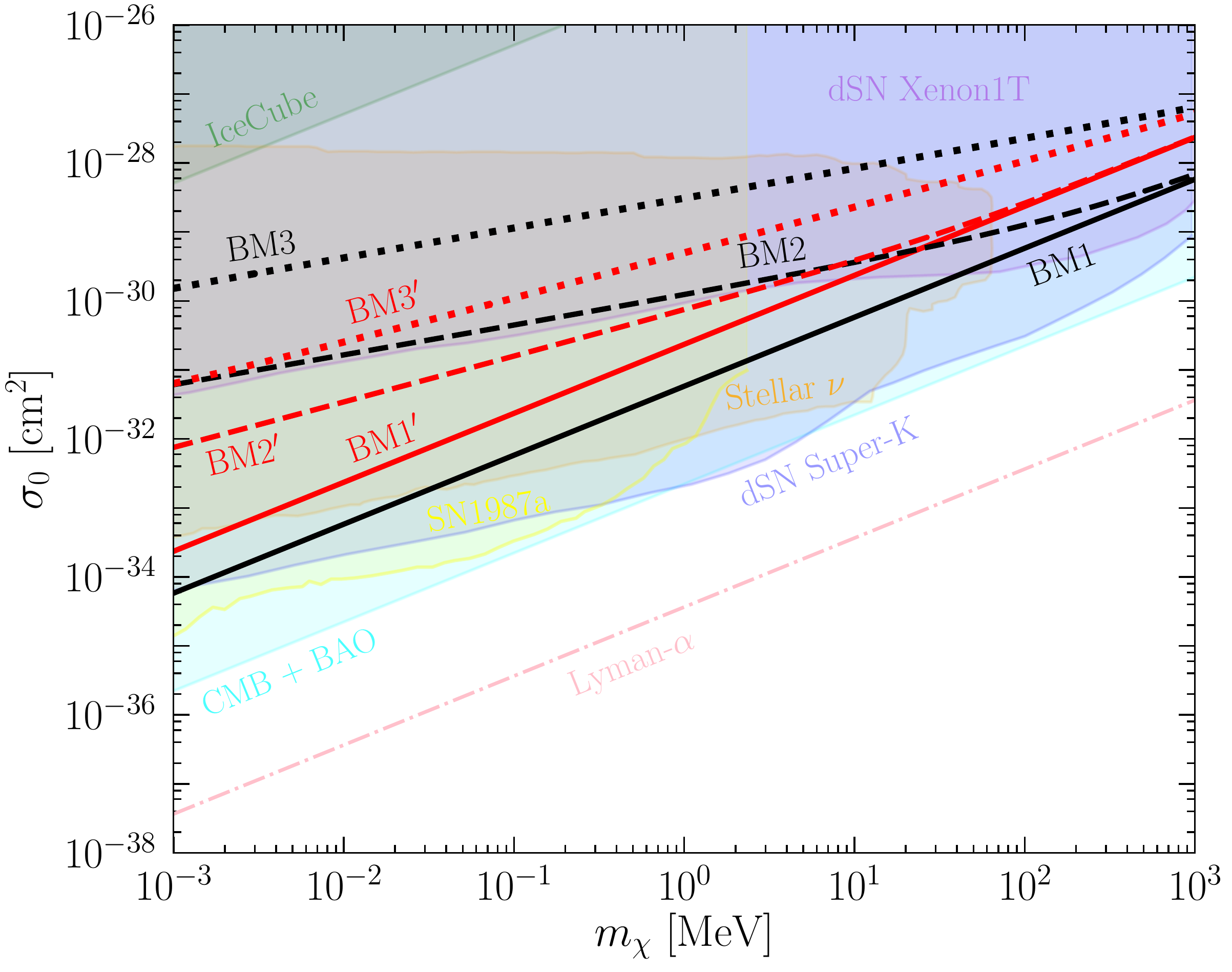}}
    \caption{\resp{$90\%$ C.L.} upper limits on the $\nu$-DM
    scattering cross section at reference energy $E_0=290\,$TeV,  for
    the six benchmark DM spike models. Previous constraints are shown
    \resp{for comparison, assuming energy-independent cross section:
    (cyan) CMB and baryon acoustic
    oscillations~\cite{Mosbech:2020ahp}; (pink) Lyman-$\alpha$
    preferred model~\cite{Hooper:2021rjc}; (dark violet, blue) diffuse
    supernova neutrinos~\cite{Ghosh:2021vkt}; (orange) stellar
    neutrinos~\cite{Jho:2021rmn}; (yellow) supernova
    SN1987A~\cite{Lin:2022dbl}; (green) IceCube bound from TXS 0506+056
    \cite{Choi:2019ixb}}.}
    \label{fig:limit}
\end{figure}

\begin{figure}[t]
   \centerline{\includegraphics[scale=0.335]{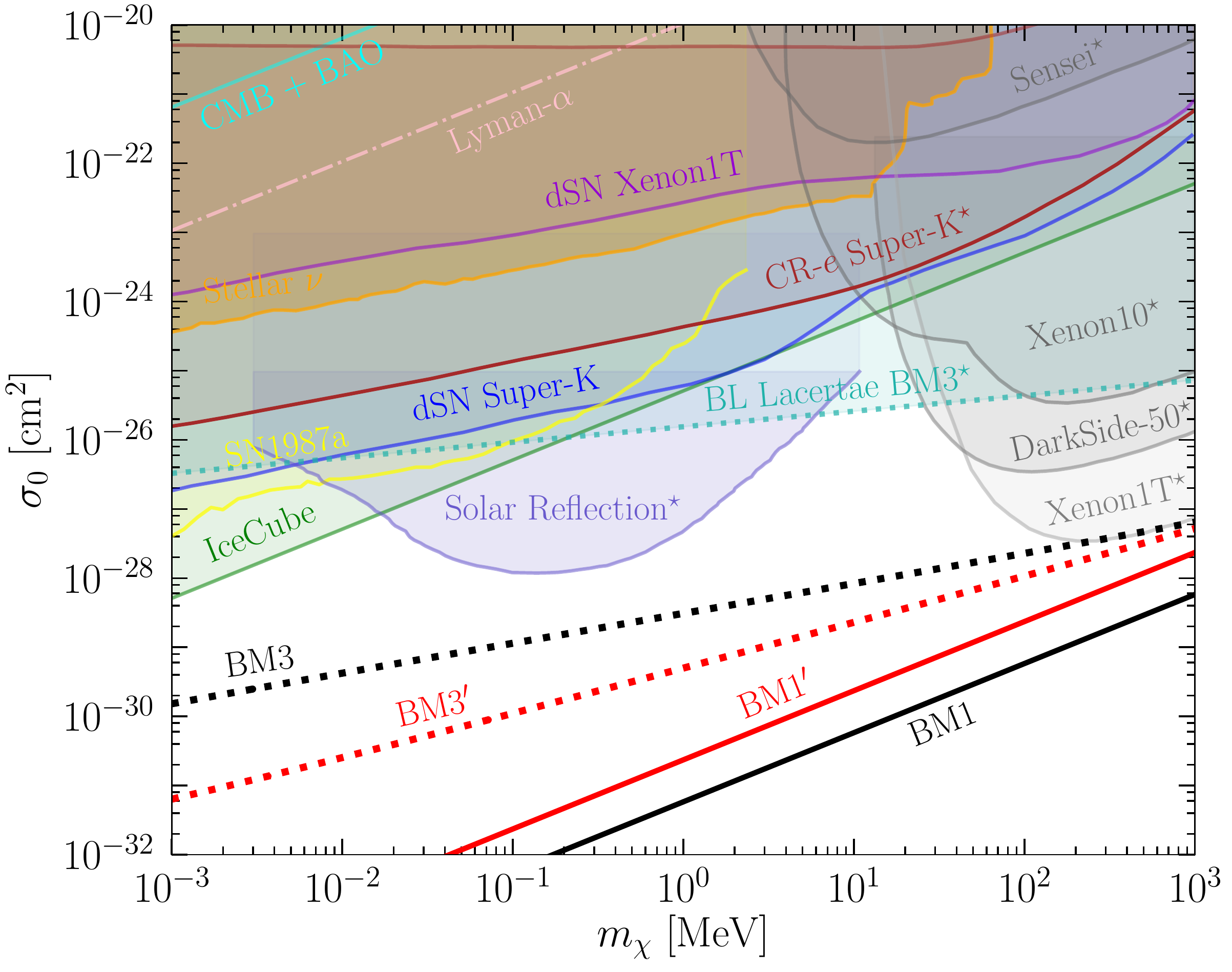}}
    \caption{
Previous constraints on $\nu$-DM and $e$-DM scattering, rescaled to
$E_0=290$\,TeV assuming $\sigma_{\nu \chi} \propto E_{\nu}$, compared to the
least \resp{(BM3, BM3$^{\prime}$)} and most restrictive \resp{(BM1,
BM1$^{\prime}$)} new limits of Fig.~\ref{fig:limit}.
    \resp{The $\nu$-DM scattering bounds are the same as in
Fig.~\ref{fig:limit},
while for $e$-DM scattering they are labelled with
$\star$ and are}: (slate blue) solar reflection~\cite{An:2017ojc}, (brown)
Super-K for DM boosted by cosmic-ray electrons, (turquoise) blazar BL Lacertae
for BM3 model~\cite{Granelli:2022ysi}, (gray) direct detection for light DM
interacting with
electrons~\cite{SENSEI:2020dpa,Essig:2017kqs,DarkSide-50:2022hin,XENON:2019gfn}.
    }  
    \label{fig:limit1}
\end{figure}

{\bf 4. Neutrino attenuation by DM.}
One can make an initial estimate for the maximum  DM-$\nu$ scattering cross section  $\sigma_{\nu\chi}$ as being inverse to the column density $\Sigma_{\chi}/m_\chi$ of the DM spike surrounding the central BH of TXS 0506+056.
To be more quantitative, 
we recompute the expected number of IceCube events from the 2017 flare
that led to the observed event, taking into account the attenuation from scattering on DM.   The analogous computation for scattering of neutrinos by galactic DM has been considered in Ref.\ \cite{Arguelles:2017atb}.  The evolution of the flux due to scattering is
described by
the cascade equation,
\be
    {d\Phi\over d\tau}(E_\nu) = -\sigma_{\nu\chi}\Phi + \int_{E_\nu}^\infty dE'_\nu\,
        {d\sigma_{\nu\chi}\over dE_\nu}({\scriptstyle E'_\nu\to E_\nu})\, \Phi(E'_\nu)\,,
        \label{cascade}
\ee  
where $\tau = \Sigma(r)/m_\chi = \int^r dr\, \rho_\chi/m_\chi$ is the accumulated
column density.
The second term represents the effect of neutrino energies being redistributed, rather than simply being lost from the beam.

To proceed, we must make an assumption about the energy dependence of the cross section. In section 6 
we will discuss particle physics models that predict $\sigma_{\nu\chi}(E_\nu)$.
  A particularly simple and well-motivated choice is linear energy dependence, 
\be
\label{eq:linsigma}
    \sigma_{\nu\chi} = {\sigma_0\,E_\nu/E_0}\,,
\ee
taking the reference energy $E_0= 290{\,\rm TeV}$ to be that of the observed 
event.  Approximating the scattering as being isotropic in the center of mass 
frame, one can show that $d\sigma_{\nu\chi}/dE_\nu = \sigma_{\nu\chi}/E'_\nu = 
\sigma_0/E_0$.  The cascade equation can be discretized, choosing equal logarithmic intervals $\Delta x$ in $x=\log_{10} (E_\nu/{\rm TeV})$.  Defining a dimensionless column density $y = (m_\chi/\Sigma_\chi)\,\tau$, it takes the form
\be
   { d\Phi_i\over dy} = A
   \left( -\hat E_i \Phi_i + \Delta x \ln 10\sum_{j=i}^N \hat E_j \Phi_j\right)
   \label{cascade2}
\ee
where $A =  (\Sigma_\chi/m_\chi)(\sigma_0/ \hat E_0)$, $\hat E_i = 10^{x_i}$ is the energy in TeV units,  $\hat E_0 = 290$ and $y\in[0,1]$. 

To solve Eq.\ (\ref{cascade2}), one can either evolve the initial condition
from $y=0$ to $y=1$ by incrementing in $y$, or use 
the algorithm presented in Ref.\ \cite{Vincent:2017svp}.  We have checked
that both methods give the same results, resulting in the \resp{90\% C.L.\
limit}
\be
    A \equiv {\Sigma_\chi\sigma_0\over m_\chi\,\hat E_0} < \resp{0.0047}
    \label{Alimit}
\ee
by demanding the number of events giving a neutrino of energy
$E_\nu\ge 290\,$TeV be greater than 0.1.  The corresponding constraints in the 
plane of $\sigma_0$ versus $m_\chi$ are plotted in Fig.\ \ref{fig:limit} for 
the six DM spike models.  The 
constraint (\ref{Alimit}) can be expressed as
\resp{$\sigma_0 < 1.4\, {m_\chi/\Sigma_\chi}$,}
in agreement with the initial estimate.
The effects of other kinds of energy dependence of $\sigma_{\nu\chi}$ are 
considered in Section 6.\resp{\footnote{If the cross section is exactly constant, the second term of the
cascade equation~\eqref{cascade} is zero and the neutrino flux is exponentially
suppressed according to $\Phi \sim \exp{(- \sigma_{\nu \chi}\,\Sigma_{\chi} /
m_{\chi})}$. The corresponding $90\%$ C.L. bound on $\sigma_{0}$ becomes $\sigma_0
\lesssim 1.7\, {m_\chi/\Sigma_\chi}$, which is very similar to the result obtained
for the case of linear energy-dependent $\sigma_{\nu \chi}$. 
}}

\resp{We find that the constraint (\ref{Alimit}) is strengthened by a factor of $\sim 4 - 10$ for
hadronic production models, like those of Refs.~\cite{Gasparyan:2021oad,Cerruti:2018tmc}, 
relative to lepto-hadronic ones.  In fact, a nonvanishing $\sigma_{\nu\chi}$
at such levels
could reduce the too-high energies predicted by hadronic models, to better
explain the IC-170922A event, but interpreted as an upper limit it is more
stringent than Eq.\ (\ref{Alimit}), hence our adoption of lepto-hadronic
models is a conservative choice.}

{\bf 5. Comparison to previous limits.}
A  model-independent signal of neutrino-DM
interactions is the suppression in the primordial density fluctuations at
temperatures $\sim 1$ eV, which would produce detectable effects in the cosmic
microwave background (CMB) and matter power
spectrum~\cite{Mangano:2006mp,Boehm:2013jpa,Bertoni:2014mva,Wilkinson:2014ksa,Mosbech:2020ahp,Hooper:2021rjc}.
For a constant scattering cross section, Ref.~\cite{Wilkinson:2014ksa} derived a
limit of $\sigma_{\nu \chi} \lesssim 10^{-36}\,(m_{\chi} /
\text{MeV})\,\,\text{cm}^2$ for massless neutrinos, which becomes weaker by about
five orders of magnitude if a neutrino mass of $\sim 0.06$ eV is properly
included~\cite{Mosbech:2020ahp}. A more recent analysis using Lyman-$\alpha$
forest data found a mild preference for DM interacting with massive neutrinos,
which requires confirmation~\cite{Hooper:2021rjc}.

Besides its effect on cosmology, DM-$\nu$ scattering can also be probed in direct detection experiments and neutrino observatories, if further assumptions about the DM interaction with either leptons or nucleons are made. 
A prominent example involves boosting DM within  our galaxy by astrophysical neutrinos such as those coming from stars~\cite{Zhang:2020nis,Jho:2021rmn}, diffuse supernovae~\cite{Farzan:2014gza,Das:2021lcr,Ghosh:2021vkt,Bardhan:2022ywd} or from supernova SN1987A~\cite{Lin:2022dbl},
leading to larger energy deposition than could occur for light
DM particles.  
Alternative ways to probe DM scattering with neutrinos is via attenuation of neutrino fluxes from supernovae~\cite{Fayet:2006sa,Mangano:2006mp} and the galactic centre~\cite{McMullen:2021ikf}, delayed neutrino propagation~\cite{Koren:2019wwi,Murase:2019xqi,Carpio:2022sml}, and through effects in the extragalactic distribution and spectra of PeV neutrinos~\cite{Davis:2015rza,Yin:2018yjn}.

Fig.~\ref{fig:limit1} shows a compilation of the most stringent bounds on $\sigma_{\nu\chi}$ after rescaling them to the common energy
scale $E_0 = 290$ TeV, assuming Eq.~\eqref{eq:linsigma}. 
Here we include also constraints on DM-electron scattering, since it is natural for neutrinos and electrons to interact with DM with the same strength, as discussed in the next section.  DM-$e$ scattering 
 can be probed in a variety of ways. It would alter the CMB anisotropies, the shape of the matter power spectrum and the abundance of Milky-Way satellites~\cite{Dvorkin:2013cea,Buen-Abad:2021mvc,Nguyen:2021cnb}, cause CMB spectral distortions~\cite{Ali-Haimoud:2015pwa,Ali-Haimoud:2021lka}, and heat or cool the gas in dwarf galaxies~\cite{Wadekar:2019mpc}. 
Similarly to the neutrino case, DM particles can be boosted by cosmic rays~\cite{Cappiello:2019qsw,Ema:2018bih,Cao:2020bwd,Xia:2020apm,Dent:2020syp,Bringmann:2018cvk,Dent:2019krz,Wang:2019jtk,Guo:2020drq,Ge:2020yuf,Jho:2020sku,Cho:2020mnc,Lei:2020mii,Guo:2020oum,Ema:2020ulo,Flambaum:2020xxo,Bell:2021xff,Feng:2021hyz,Wang:2021nbf,Xia:2021vbz,Bramante:2021dyx}, particles in the solar interior~\cite{An:2017ojc} or in the relativistic jets of blazars~\cite{Wang:2021jic,Granelli:2022ysi} and be directly detected. Standard direct detection constraints on light DM particles can apply~\cite{Essig:2012yx,Essig:2017kqs,SuperCDMS:2018mne,SuperCDMS:2020ymb,DarkSide:2018ppu,DarkSide-50:2022hin,Crisler:2018gci,SENSEI:2019ibb,SENSEI:2020dpa,XENON:2019gfn,XENON:2021nad,DAMIC:2019dcn,EDELWEISS:2020fxc,PandaX-II:2021nsg}.
DM-electron scattering can alter the cosmic ray spectrum~\cite{Cappiello:2018hsu}, and \resp{potentially} heat neutron stars~\cite{Bell:2019pyc,Bell:2020jou,Bell:2020lmm} \resp{and} white dwarfs~\cite{Bell:2021fye}.

The new blazar limits on $\sigma_{\nu\chi}$ shown in  Fig.\ \ref{fig:limit1},
assuming $\sigma_{\nu\chi}\propto E_{\nu}$, are several orders of magnitude stronger than existing ones for sub-GeV DM, when the latter are rescaled to the blazar neutrino energy.  In the case of light mediators that could lead to a constant-in-energy cross section, we lose this advantage, as shown in 
Fig.\ \ref{fig:limit}.

{\bf 6. Particle physics models.}
The simplest models for DM-$\nu$ scattering involve the exchange of
a vector boson $Z'$ between DM and 
neutrinos. 
We 
assume coupling $g _\nu$ to all
flavors of neutrinos, and coupling $g_\chi$ to DM,
taken to be a complex scalar; by dimensional analysis, the results are expected to be 
insensitive to  the spin of the DM.  
(Exact expressions for $\sigma(E)$ in various models can be found in the appendix of Ref. \cite{Arguelles:2017atb}.) At energies $E_{\nu} \gg m_\chi$,
the cross section goes as
\be 
\sigma_{\nu\chi} \cong { g_{\nu}^2 g_{\chi}^2 \over 4\pi\, m_{Z'}^2}\left[1 - {m_{Z'}^2\over s}
    \ln \left(1 + {s\over m_{Z'}^2}\right)\right]\,,
\ee
where $s \cong 2 m_\chi E_\nu$, equally for scattering of neutrinos or muons on DM.  
For $m_{Z'}^2 >  m_\chi E_\nu \gtrsim 1\,$GeV$^2$ (considering $m_\chi$ as low as 1 keV), $\sigma_{\nu\chi}$ rises linearly with $E_\nu$
by expanding the logarithm to second order in $s/m_{Z'}^2$, while for $E_\nu\gg
m_{Z'}^2/m_\chi$, $\sigma_{\nu\chi}$ saturates to a constant value.
The corresponding differential cross section that appears in the second term of the cascade
equation (\ref{cascade}) is
\be
    {d\sigma_{\nu\chi}\over dE_{\nu}}({\scriptstyle E_{\nu}'\to E_{\nu}}) = {(g_{\nu}^2 g_{\chi}^2 /4\pi) (m_\chi\,E_{\nu}/E_{\nu}')  \over
      (m_{Z'}^2 + 2 m_\chi(E_{\nu}'-E_{\nu}))^2}\,.
      \label{dsdE}
\ee

\begin{figure}[t]
   \centerline{\includegraphics[scale=0.27]{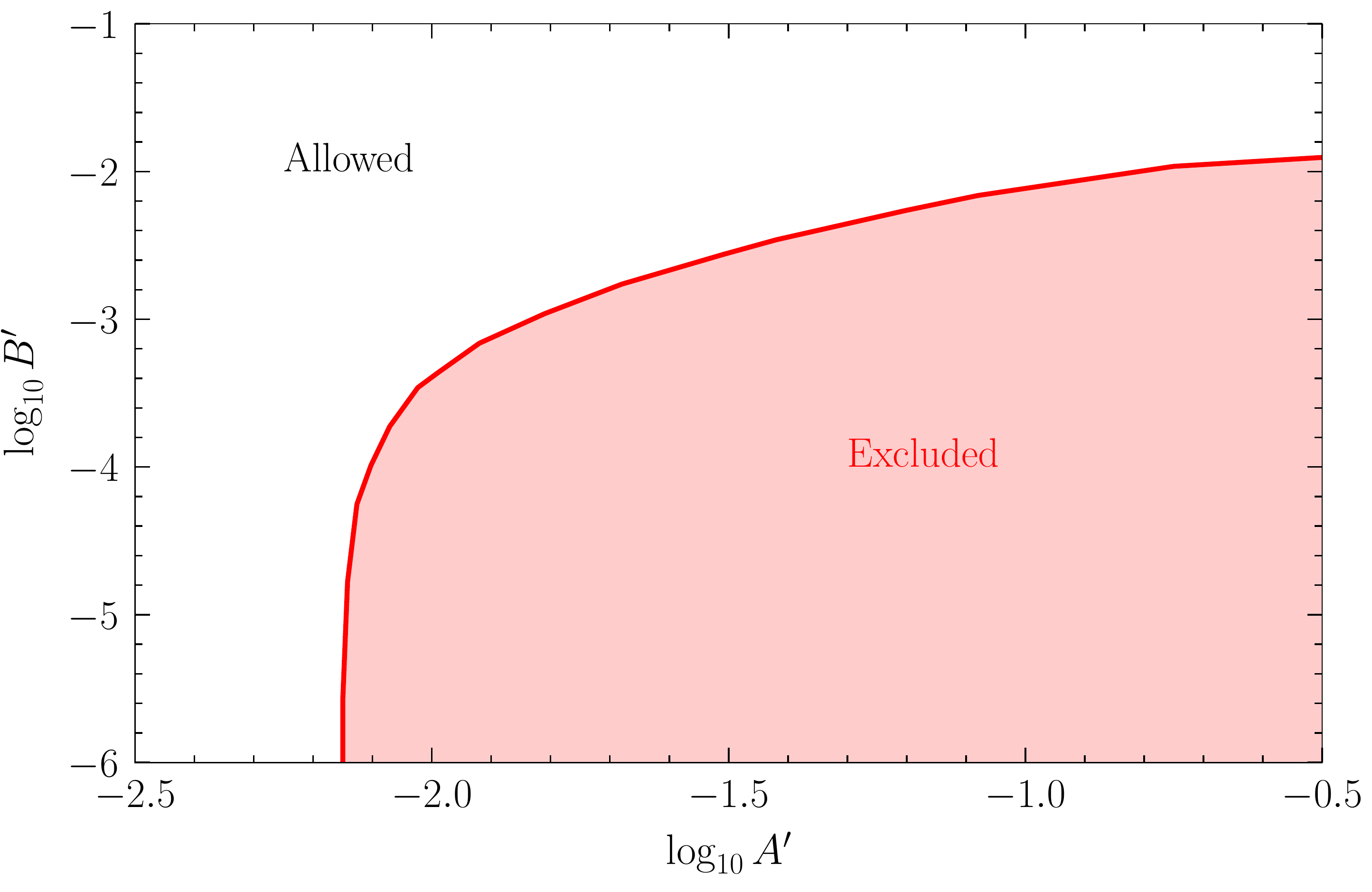}}
    \caption{Constraint on the dimensionless parameters defined in 
    Eq.\ (\ref{ApBp}) in the model with a $Z'$ mediator.}  
    \label{fig:limit2}
\end{figure}

\begin{figure}[t]
   \includegraphics[scale=0.27]{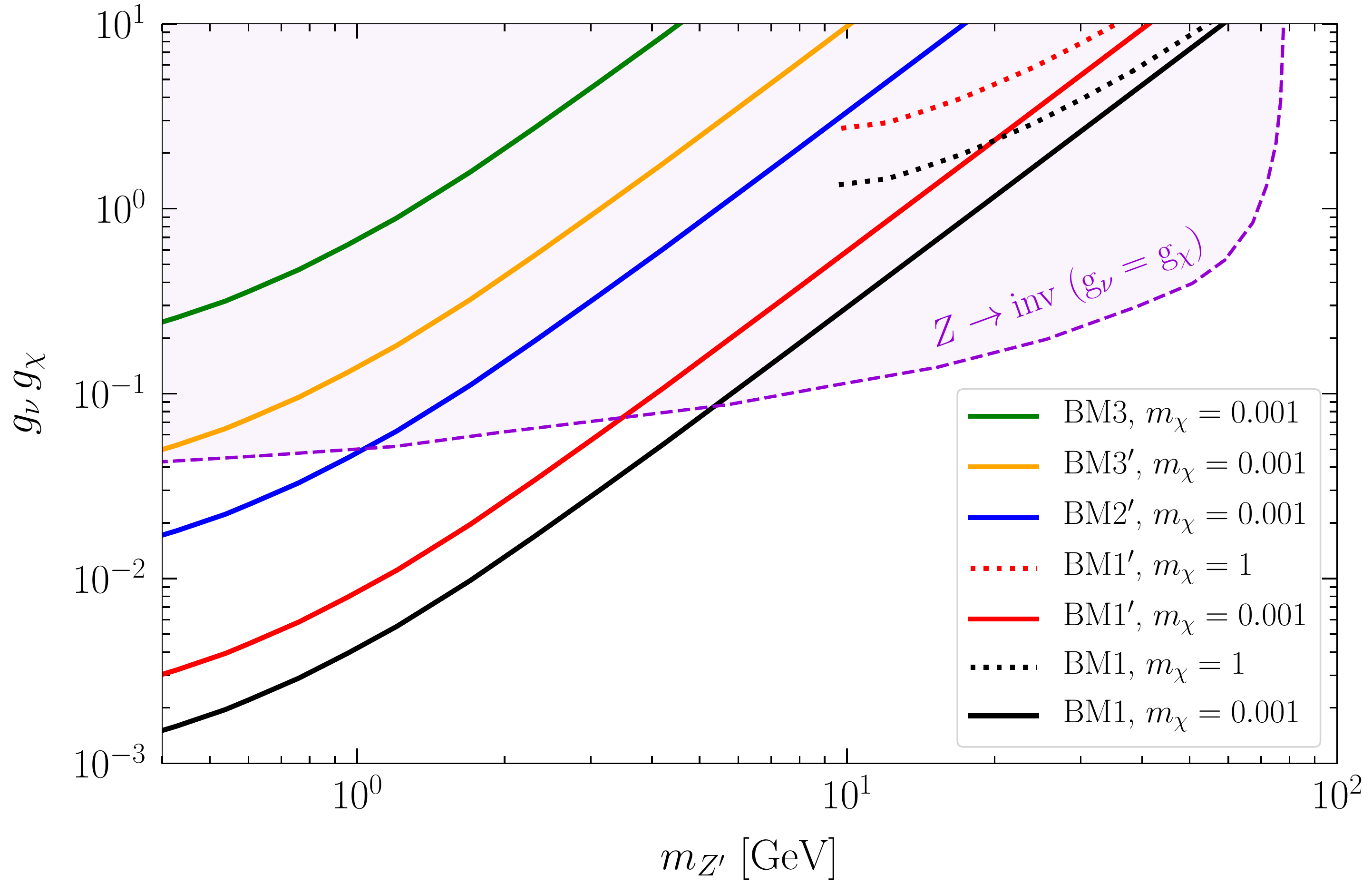}
    \caption{Upper limit on \resp{the product of the couplings $g_{\nu} g_{\chi}$} versus $m_{Z'}$ in the vector boson mediator model,
    for several choices of DM spike model and mass $m_\chi$, indicated in MeV units.
    \resp{Laboratory bound from $Z\to 4\nu$ \cite{Bilenky:1992xn,Berryman:2022hds} 
	is shown for the case $g_\chi=g_\nu$.}}
    \label{fig:g2const}
\end{figure}

This model is similar to that in Eq.\ (\ref{eq:linsigma}) in having $\sigma_{\nu\chi}\propto E_\nu$ at low energy, but it is physically distinct because
the differential scattering implied by (\ref{dsdE}) is not isotropic.  
One can show that its behavior in the cascade equation is determined by just
two (dimensionless) parameters, that we take to be
\be
    A' = {g_{\nu}^2 g_{\chi}^2\, \Sigma_\chi\cdot
(1\,{\rm TeV})\over 4\pi\,m_{Z'}^4},\quad
 B'\equiv {m_\chi \cdot
(1\,{\rm TeV})\over m_{Z'}^2}\,.
\label{ApBp}
\ee 
With this choice, $A'$ plays the same role of $A$ in Eq.\ 
(\ref{cascade2}) in the low-energy regime where $\sigma_{\nu\chi} \sim g_{\nu}^2 g_{\chi}^2 m_\chi E_{\nu}/(4\pi m_{Z'}^4)$.
By solving the cascade equation on a grid of values in the $A'$-$B'$ plane,
again demanding at least 0.1 predicted IceCube events above 290 TeV,
we obtain the constraint shown in Fig.\ \ref{fig:limit2}.   
We translate the $A'$ versus $B'$ bound into the microscopic model parameters, 
$g_\nu g_\chi$ versus $m_{Z'}$ in Fig.\ \ref{fig:g2const}, for some choices of the DM spike models and DM masses.  
\resp{For comparison, the most stringent related constraint from $Z\to 4\nu$ is
also shown \cite{Bilenky:1992xn,Berryman:2022hds}, for the case that $g_\chi=g_\nu$.}

In a realistic model, $Z'$ should couple not only to neutrinos, but to charged leptons in the SU(2)$_L$
doublets, and to baryons so that the theory is anomaly-free.  This leads to numerous further constraints in the parameter space of $g_\nu$ versus $m_{Z'}$, which are beyond the scope of the present work.  We will consider this aspect in an upcoming paper
\cite{inprogress}.


{\bf 7. Summary and conclusions.}
It is not disputed that dark matter accumulates in the vicinity of supermassive black holes that power active galactic nuclei, but there are  significant
uncertainties from astrophysics\resp{, including the initial neutrino flux and the location along the jet where neutrinos are likely to be produced, 
and from the density profile of the DM spike and the effective DM annihilation cross section.} \resp{Despite these uncertainties,} we
find strong \resp{and conservative} constraints on the elastic scattering cross section $\sigma_{\nu\chi}$ for DM-neutrino scattering, so long as the
IceCube event IC-170922A indeed came from the blazar TXS 0506+056 during its 2017 flare, as is widely believed.  

Since the single event has a unique neutrino energy $E_0$, our constraint
applies to $\sigma_{\nu\chi}$ at that energy.  A natural hypothesis is that such interactions arise from exchange of a massive mediator, which leads to the prediction of linear energy dependence, $\sigma_{\nu\chi} = \sigma_0\, E_{\nu}/E_0$ at sufficiently low energies.  Under that assumption, we compared our limit to previous ones in the literature, which are set at much lower energies. 

Even in the least optimistic case (models BM3-BM3$'$), our limits improve on the existing ones  by several orders of magnitude, \resp{if rescaled to $E_{0}$}, for sub-GeV DM masses \resp{(see Fig.~\ref{fig:limit1})}. The stronger of our constraints (BM1-BM$1'$) are likely to be applicable in the case of asymmetric DM,
where the effective annihilation cross section is essentially zero, due to the
negligible proportion of a symmetric component that is necessary to have
annihilation. Our constraints are weakened if the mediator mass is sufficiently
small, which causes the cross section to stop rising with energy at a scale of order
$m_{Z'}^2/m_\chi$, becoming constant at higher energies, and thereby reducing the
leverage of our bound coming from the 290 TeV scale \resp{(see
Fig.~\ref{fig:limit})}.

A further natural assumption, motivated  by SU(2)$_L$ gauge symmetry in the standard model, is that charged leptons should have an equal cross section with DM relative to neutrinos, allowing us to compare to existing
electron-DM scattering constraints.  Here too our constraints improve on previous limits, for linearly rising cross sections.

We look forward to future observations by neutrino telescopes that may confirm the multimessenger signals from blazars, and perhaps lead to refined constraints on lepton-DM scattering.

{\bf Acknowledgments.} We thank Sargis Gasparyan, Rebecca Leane, Kohta Murase, Foteini Oikonomou, Paolo Padovani, Ken Ragan, Aaron Vincent and Jin-Wei Wang for helpful correspondence.
This work was supported by the Natural Sciences and Engineering Research Council (NSERC) of Canada.

\bibliographystyle{utphys}
\bibliography{ref2.bib}

\providecommand{\href}[2]{#2}\begingroup\raggedright\begin{thebibliography}{100}

\bibitem{Wang:2021jic}
J.-W. Wang, A.~Granelli, and P.~Ullio, ``{Direct Detection Constraints on
  Blazar-Boosted Dark Matter},''
  \href{http://dx.doi.org/10.1103/PhysRevLett.128.221104}{{\em Phys. Rev.
  Lett.} {\bfseries 128} (2022) 221104},
  \href{http://arxiv.org/abs/2111.13644}{{\ttfamily arXiv:2111.13644
  [astro-ph.HE]}}.

\bibitem{Ghosh:2021vkt}
D.~Ghosh, A.~Guha, and D.~Sachdeva, ``{Exclusion limits on dark matter-neutrino
  scattering cross section},''
  \href{http://dx.doi.org/10.1103/PhysRevD.105.103029}{{\em Phys. Rev. D}
  {\bfseries 105} no.~10, (2022) 103029},
  \href{http://arxiv.org/abs/2110.00025}{{\ttfamily arXiv:2110.00025
  [hep-ph]}}.

\bibitem{Granelli:2022ysi}
A.~Granelli, P.~Ullio, and J.-W. Wang, ``{Blazar-boosted dark matter at
  Super-Kamiokande},''
  \href{http://dx.doi.org/10.1088/1475-7516/2022/07/013}{{\em JCAP} {\bfseries
  07} no.~07, (2022) 013}, \href{http://arxiv.org/abs/2202.07598}{{\ttfamily
  arXiv:2202.07598 [astro-ph.HE]}}.

\bibitem{IceCube:2018dnn}
{\bfseries IceCube, Fermi-LAT, MAGIC, AGILE, ASAS-SN, HAWC, H.E.S.S., INTEGRAL,
  Kanata, Kiso, Kapteyn, Liverpool Telescope, Subaru, Swift NuSTAR, VERITAS,
  VLA/17B-403} Collaboration, M.~G. Aartsen {\em et~al.}, ``{Multimessenger
  observations of a flaring blazar coincident with high-energy neutrino
  IceCube-170922A},'' \href{http://dx.doi.org/10.1126/science.aat1378}{{\em
  Science} {\bfseries 361} no.~6398, (2018) eaat1378},
  \href{http://arxiv.org/abs/1807.08816}{{\ttfamily arXiv:1807.08816
  [astro-ph.HE]}}.

\bibitem{Choi:2019ixb}
K.-Y. Choi, J.~Kim, and C.~Rott, ``{Constraining dark matter-neutrino
  interactions with IceCube-170922A},''
  \href{http://dx.doi.org/10.1103/PhysRevD.99.083018}{{\em Phys. Rev. D}
  {\bfseries 99} no.~8, (2019) 083018},
  \href{http://arxiv.org/abs/1903.03302}{{\ttfamily arXiv:1903.03302
  [astro-ph.CO]}}.

\bibitem{IceCube:2018cha}
{\bfseries IceCube} Collaboration, M.~G. Aartsen {\em et~al.}, ``{Neutrino
  emission from the direction of the blazar TXS 0506+056 prior to the
  IceCube-170922A alert},''
  \href{http://dx.doi.org/10.1126/science.aat2890}{{\em Science} {\bfseries
  361} no.~6398, (2018) 147--151},
  \href{http://arxiv.org/abs/1807.08794}{{\ttfamily arXiv:1807.08794
  [astro-ph.HE]}}.

\bibitem{Keivani:2018rnh}
A.~Keivani {\em et~al.}, ``{A Multimessenger Picture of the Flaring Blazar TXS
  0506+056: implications for High-Energy Neutrino Emission and Cosmic Ray
  Acceleration},'' \href{http://dx.doi.org/10.3847/1538-4357/aad59a}{{\em
  Astrophys. J.} {\bfseries 864} no.~1, (2018) 84},
  \href{http://arxiv.org/abs/1807.04537}{{\ttfamily arXiv:1807.04537
  [astro-ph.HE]}}.

\bibitem{Murase:2018iyl}
K.~Murase, F.~Oikonomou, and M.~Petropoulou, ``{Blazar Flares as an Origin of
  High-Energy Cosmic Neutrinos?},''
  \href{http://dx.doi.org/10.3847/1538-4357/aada00}{{\em Astrophys. J.}
  {\bfseries 865} no.~2, (2018) 124},
  \href{http://arxiv.org/abs/1807.04748}{{\ttfamily arXiv:1807.04748
  [astro-ph.HE]}}.

\bibitem{Reimer:2018vvw}
A.~Reimer, M.~Boettcher, and S.~Buson, ``{Cascading Constraints from
  Neutrino-emitting Blazars: The Case of TXS 0506+056},''
  \href{http://dx.doi.org/10.3847/1538-4357/ab2bff}{{\em Astrophys. J.}
  {\bfseries 881} no.~1, (2019) 46},
  \href{http://arxiv.org/abs/1812.05654}{{\ttfamily arXiv:1812.05654
  [astro-ph.HE]}}. [Erratum: Astrophys.J. 899, 168 (2020)].

\bibitem{Rodrigues:2018tku}
X.~Rodrigues, S.~Gao, A.~Fedynitch, A.~Palladino, and W.~Winter,
  ``{Leptohadronic Blazar Models Applied to the 2014\textendash{}2015 Flare of
  TXS 0506+056},'' \href{http://dx.doi.org/10.3847/2041-8213/ab1267}{{\em
  Astrophys. J. Lett.} {\bfseries 874} no.~2, (2019) L29},
  \href{http://arxiv.org/abs/1812.05939}{{\ttfamily arXiv:1812.05939
  [astro-ph.HE]}}.

\bibitem{Petropoulou:2019zqp}
M.~Petropoulou {\em et~al.}, ``{Multi-Epoch Modeling of TXS 0506+056 and
  Implications for Long-Term High-Energy Neutrino Emission},''
  \href{http://dx.doi.org/10.3847/1538-4357/ab76d0}{{\em Astrophys. J.}
  {\bfseries 891} (2020) 115},
  \href{http://arxiv.org/abs/1911.04010}{{\ttfamily arXiv:1911.04010
  [astro-ph.HE]}}.

\bibitem{Gasparyan:2021oad}
S.~Gasparyan, D.~B\'egu\'e, and N.~Sahakyan, ``{Time-dependent lepto-hadronic
  modelling of the emission from blazar jets with SOPRANO: the case of TXS
  0506~+~056, 3HSP J095507.9~+~355101, and 3C 279},''
  \href{http://dx.doi.org/10.1093/mnras/stab2688}{{\em Mon. Not. Roy. Astron.
  Soc.} {\bfseries 509} no.~2, (2021) 2102--2121},
  \href{http://arxiv.org/abs/2110.01549}{{\ttfamily arXiv:2110.01549
  [astro-ph.HE]}}.

\bibitem{Rodrigues:2020fbu}
X.~Rodrigues, S.~Garrappa, S.~Gao, V.~S. Paliya, A.~Franckowiak, and W.~Winter,
  ``{Multiwavelength and Neutrino Emission from Blazar PKS 1502 + 106},''
  \href{http://dx.doi.org/10.3847/1538-4357/abe87b}{{\em Astrophys. J.}
  {\bfseries 912} no.~1, (2021) 54},
  \href{http://arxiv.org/abs/2009.04026}{{\ttfamily arXiv:2009.04026
  [astro-ph.HE]}}.

\bibitem{Giommi:2020viy}
P.~Giommi, P.~Padovani, F.~Oikonomou, T.~Glauch, S.~Paiano, and E.~Resconi,
  ``{3HSP J095507.9+355101: a flaring extreme blazar coincident in space and
  time with IceCube-200107A},''
  \href{http://dx.doi.org/10.1051/0004-6361/202038423}{{\em Astron. Astrophys.}
  {\bfseries 640} (2020) L4}, \href{http://arxiv.org/abs/2003.06405}{{\ttfamily
  arXiv:2003.06405 [astro-ph.HE]}}.

\bibitem{Fermi-LAT:2019hte}
{\bfseries Fermi-LAT, ASAS-SN, IceCube} Collaboration, S.~Garrappa {\em
  et~al.}, ``{Investigation of two Fermi-LAT gamma-ray blazars coincident with
  high-energy neutrinos detected by IceCube},''
  \href{http://dx.doi.org/10.3847/1538-4357/ab2ada}{{\em Astrophys. J.}
  {\bfseries 880} no.~2, (2019) 880:103},
  \href{http://arxiv.org/abs/1901.10806}{{\ttfamily arXiv:1901.10806
  [astro-ph.HE]}}.

\bibitem{Kadler:2016ygj}
M.~Kadler {\em et~al.}, ``{Coincidence of a high-fluence blazar outburst with a
  PeV-energy neutrino event},'' \href{http://dx.doi.org/10.1038/NPHYS3715}{{\em
  Nature Phys.} {\bfseries 12} no.~8, (2016) 807--814},
  \href{http://arxiv.org/abs/1602.02012}{{\ttfamily arXiv:1602.02012
  [astro-ph.HE]}}.

\bibitem{Sahakyan:2022nbz}
N.~Sahakyan, P.~Giommi, P.~Padovani, M.~Petropoulou, D.~B\'egu\'e, B.~Boccardi,
  and S.~Gasparyan, ``{A multi-messenger study of the blazar PKS 0735+178: a
  new major neutrino source candidate},''
  \href{http://arxiv.org/abs/2204.05060}{{\ttfamily arXiv:2204.05060
  [astro-ph.HE]}}.

\bibitem{IceCube:2021slf}
{\bfseries IceCube} Collaboration, R.~Abbasi {\em et~al.}, ``{Search for
  Multi-flare Neutrino Emissions in 10 yr of IceCube Data from a Catalog of
  Sources},'' \href{http://dx.doi.org/10.3847/2041-8213/ac2c7b}{{\em Astrophys.
  J. Lett.} {\bfseries 920} no.~2, (2021) L45},
  \href{http://arxiv.org/abs/2109.05818}{{\ttfamily arXiv:2109.05818
  [astro-ph.HE]}}.

\bibitem{Giommi:2020hbx}
P.~Giommi, T.~Glauch, P.~Padovani, E.~Resconi, A.~Turcati, and Y.~L. Chang,
  ``{Dissecting the regions around IceCube high-energy neutrinos: growing
  evidence for the blazar connection},''
  \href{http://dx.doi.org/10.1093/mnras/staa2082}{{\em Mon. Not. Roy. Astron.
  Soc.} {\bfseries 497} no.~1, (2020) 865--878},
  \href{http://arxiv.org/abs/2001.09355}{{\ttfamily arXiv:2001.09355
  [astro-ph.HE]}}.

\bibitem{Franckowiak:2020qrq}
A.~Franckowiak {\em et~al.}, ``{Patterns in the Multiwavelength Behavior of
  Candidate Neutrino Blazars},''
  \href{http://dx.doi.org/10.3847/1538-4357/ab8307}{{\em Astrophys. J.}
  {\bfseries 893} no.~2, (2020) 162},
  \href{http://arxiv.org/abs/2001.10232}{{\ttfamily arXiv:2001.10232
  [astro-ph.HE]}}.

\bibitem{Mucke:2000rn}
A.~Mucke and R.~J. Protheroe, ``{A Proton synchrotron blazar model for flaring
  in Markarian 501},''
  \href{http://dx.doi.org/10.1016/S0927-6505(00)00141-9}{{\em Astropart. Phys.}
  {\bfseries 15} (2001) 121--136},
  \href{http://arxiv.org/abs/astro-ph/0004052}{{\ttfamily
  arXiv:astro-ph/0004052}}.

\bibitem{Cerruti:2018tmc}
M.~Cerruti, A.~Zech, C.~Boisson, G.~Emery, S.~Inoue, and J.~P. Lenain,
  ``{Leptohadronic single-zone models for the electromagnetic and neutrino
  emission of TXS 0506+056},''
  \href{http://dx.doi.org/10.1093/mnrasl/sly210}{{\em Mon. Not. Roy. Astron.
  Soc.} {\bfseries 483} no.~1, (2019) L12--L16},
  \href{http://arxiv.org/abs/1807.04335}{{\ttfamily arXiv:1807.04335
  [astro-ph.HE]}}. [Erratum: Mon.Not.Roy.Astron.Soc. 502, L21--L22 (2021)].

\bibitem{Gao:2018mnu}
S.~Gao, A.~Fedynitch, W.~Winter, and M.~Pohl, ``{Modelling the coincident
  observation of a high-energy neutrino and a bright blazar flare},''
  \href{http://dx.doi.org/10.1038/s41550-018-0610-1}{{\em Nature Astron.}
  {\bfseries 3} no.~1, (2019) 88--92},
  \href{http://arxiv.org/abs/1807.04275}{{\ttfamily arXiv:1807.04275
  [astro-ph.HE]}}.

\bibitem{icecube-data}
{\bfseries IceCube} Collaboration.
  \url{https://icecube.wisc.edu/data-releases/2018/07/}.

\bibitem{Gondolo:1999ef}
P.~Gondolo and J.~Silk, ``{Dark matter annihilation at the galactic center},''
  \href{http://dx.doi.org/10.1103/PhysRevLett.83.1719}{{\em Phys. Rev. Lett.}
  {\bfseries 83} (1999) 1719--1722},
  \href{http://arxiv.org/abs/astro-ph/9906391}{{\ttfamily
  arXiv:astro-ph/9906391}}.

\bibitem{Ullio:2001fb}
P.~Ullio, H.~Zhao, and M.~Kamionkowski, ``{A Dark matter spike at the galactic
  center?},'' \href{http://dx.doi.org/10.1103/PhysRevD.64.043504}{{\em Phys.
  Rev. D} {\bfseries 64} (2001) 043504},
  \href{http://arxiv.org/abs/astro-ph/0101481}{{\ttfamily
  arXiv:astro-ph/0101481}}.

\bibitem{Gorchtein:2010xa}
M.~Gorchtein, S.~Profumo, and L.~Ubaldi, ``{Probing Dark Matter with AGN
  Jets},'' \href{http://dx.doi.org/10.1103/PhysRevD.82.083514}{{\em Phys. Rev.
  D} {\bfseries 82} (2010) 083514},
  \href{http://arxiv.org/abs/1008.2230}{{\ttfamily arXiv:1008.2230
  [astro-ph.HE]}}. [Erratum: Phys.Rev.D 84, 069903 (2011)].

\bibitem{Padovani:2019xcv}
P.~Padovani, F.~Oikonomou, M.~Petropoulou, P.~Giommi, and E.~Resconi, ``{TXS
  0506+056, the first cosmic neutrino source, is not a BL Lac},''
  \href{http://dx.doi.org/10.1093/mnrasl/slz011}{{\em Mon. Not. Roy. Astron.
  Soc.} {\bfseries 484} no.~1, (2019) L104--L108},
  \href{http://arxiv.org/abs/1901.06998}{{\ttfamily arXiv:1901.06998
  [astro-ph.HE]}}.

\bibitem{Gnedin:2003rj}
O.~Y. Gnedin and J.~R. Primack, ``{Dark Matter Profile in the Galactic
  Center},'' \href{http://dx.doi.org/10.1103/PhysRevLett.93.061302}{{\em Phys.
  Rev. Lett.} {\bfseries 93} (2004) 061302},
  \href{http://arxiv.org/abs/astro-ph/0308385}{{\ttfamily
  arXiv:astro-ph/0308385}}.

\bibitem{Mosbech:2020ahp}
M.~R. Mosbech, C.~Boehm, S.~Hannestad, O.~Mena, J.~Stadler, and Y.~Y.~Y. Wong,
  ``{The full Boltzmann hierarchy for dark matter-massive neutrino
  interactions},'' \href{http://dx.doi.org/10.1088/1475-7516/2021/03/066}{{\em
  JCAP} {\bfseries 03} (2021) 066},
  \href{http://arxiv.org/abs/2011.04206}{{\ttfamily arXiv:2011.04206
  [astro-ph.CO]}}.

\bibitem{Hooper:2021rjc}
D.~C. Hooper and M.~Lucca, ``{Hints of dark matter-neutrino interactions in
  Lyman-\ensuremath{\alpha} data},''
  \href{http://dx.doi.org/10.1103/PhysRevD.105.103504}{{\em Phys. Rev. D}
  {\bfseries 105} no.~10, (2022) 103504},
  \href{http://arxiv.org/abs/2110.04024}{{\ttfamily arXiv:2110.04024
  [astro-ph.CO]}}.

\bibitem{Jho:2021rmn}
Y.~Jho, J.-C. Park, S.~C. Park, and P.-Y. Tseng, ``{Cosmic-Neutrino-Boosted
  Dark Matter ($\nu$BDM)},'' \href{http://arxiv.org/abs/2101.11262}{{\ttfamily
  arXiv:2101.11262 [hep-ph]}}.

\bibitem{Lin:2022dbl}
Y.-H. Lin, W.-H. Wu, M.-R. Wu, and H.~T.-K. Wong, ``{Searching for Afterglow:
  Light Dark Matter boosted by Supernova Neutrinos},''
  \href{http://arxiv.org/abs/2206.06864}{{\ttfamily arXiv:2206.06864
  [hep-ph]}}.

\bibitem{An:2017ojc}
H.~An, M.~Pospelov, J.~Pradler, and A.~Ritz, ``{Directly Detecting MeV-scale
  Dark Matter via Solar Reflection},''
  \href{http://dx.doi.org/10.1103/PhysRevLett.120.141801}{{\em Phys. Rev.
  Lett.} {\bfseries 120} no.~14, (2018) 141801},
  \href{http://arxiv.org/abs/1708.03642}{{\ttfamily arXiv:1708.03642
  [hep-ph]}}. [Erratum: Phys.Rev.Lett. 121, 259903 (2018)].

\bibitem{SENSEI:2020dpa}
{\bfseries SENSEI} Collaboration, L.~Barak {\em et~al.}, ``{SENSEI:
  Direct-Detection Results on sub-GeV Dark Matter from a New Skipper-CCD},''
  \href{http://dx.doi.org/10.1103/PhysRevLett.125.171802}{{\em Phys. Rev.
  Lett.} {\bfseries 125} no.~17, (2020) 171802},
  \href{http://arxiv.org/abs/2004.11378}{{\ttfamily arXiv:2004.11378
  [astro-ph.CO]}}.

\bibitem{Essig:2017kqs}
R.~Essig, T.~Volansky, and T.-T. Yu, ``{New Constraints and Prospects for
  sub-GeV Dark Matter Scattering off Electrons in Xenon},''
  \href{http://dx.doi.org/10.1103/PhysRevD.96.043017}{{\em Phys. Rev. D}
  {\bfseries 96} no.~4, (2017) 043017},
  \href{http://arxiv.org/abs/1703.00910}{{\ttfamily arXiv:1703.00910
  [hep-ph]}}.

\bibitem{DarkSide-50:2022hin}
{\bfseries DarkSide-50} Collaboration, P.~Agnes {\em et~al.}, ``{Search for
  dark matter particle interactions with electron final states with
  DarkSide-50},'' \href{http://arxiv.org/abs/2207.11968}{{\ttfamily
  arXiv:2207.11968 [hep-ex]}}.

\bibitem{XENON:2019gfn}
{\bfseries XENON} Collaboration, E.~Aprile {\em et~al.}, ``{Light Dark Matter
  Search with Ionization Signals in XENON1T},''
  \href{http://dx.doi.org/10.1103/PhysRevLett.123.251801}{{\em Phys. Rev.
  Lett.} {\bfseries 123} no.~25, (2019) 251801},
  \href{http://arxiv.org/abs/1907.11485}{{\ttfamily arXiv:1907.11485
  [hep-ex]}}.

\bibitem{Arguelles:2017atb}
C.~A. Arg\"uelles, A.~Kheirandish, and A.~C. Vincent, ``{Imaging Galactic Dark
  Matter with High-Energy Cosmic Neutrinos},''
  \href{http://dx.doi.org/10.1103/PhysRevLett.119.201801}{{\em Phys. Rev.
  Lett.} {\bfseries 119} no.~20, (2017) 201801},
  \href{http://arxiv.org/abs/1703.00451}{{\ttfamily arXiv:1703.00451
  [hep-ph]}}.

\bibitem{Vincent:2017svp}
A.~C. Vincent, C.~A. Arg\"uelles, and A.~Kheirandish, ``{High-energy neutrino
  attenuation in the Earth and its associated uncertainties},''
  \href{http://dx.doi.org/10.1088/1475-7516/2017/11/012}{{\em JCAP} {\bfseries
  11} (2017) 012}, \href{http://arxiv.org/abs/1706.09895}{{\ttfamily
  arXiv:1706.09895 [hep-ph]}}.

\bibitem{Mangano:2006mp}
G.~Mangano, A.~Melchiorri, P.~Serra, A.~Cooray, and M.~Kamionkowski,
  ``{Cosmological bounds on dark matter-neutrino interactions},''
  \href{http://dx.doi.org/10.1103/PhysRevD.74.043517}{{\em Phys. Rev. D}
  {\bfseries 74} (2006) 043517},
  \href{http://arxiv.org/abs/astro-ph/0606190}{{\ttfamily
  arXiv:astro-ph/0606190}}.

\bibitem{Boehm:2013jpa}
C.~Boehm, M.~J. Dolan, and C.~McCabe, ``{A Lower Bound on the Mass of Cold
  Thermal Dark Matter from Planck},''
  \href{http://dx.doi.org/10.1088/1475-7516/2013/08/041}{{\em JCAP} {\bfseries
  08} (2013) 041}, \href{http://arxiv.org/abs/1303.6270}{{\ttfamily
  arXiv:1303.6270 [hep-ph]}}.

\bibitem{Bertoni:2014mva}
B.~Bertoni, S.~Ipek, D.~McKeen, and A.~E. Nelson, ``{Constraints and
  consequences of reducing small scale structure via large dark matter-neutrino
  interactions},'' \href{http://dx.doi.org/10.1007/JHEP04(2015)170}{{\em JHEP}
  {\bfseries 04} (2015) 170}, \href{http://arxiv.org/abs/1412.3113}{{\ttfamily
  arXiv:1412.3113 [hep-ph]}}.

\bibitem{Wilkinson:2014ksa}
R.~J. Wilkinson, C.~Boehm, and J.~Lesgourgues, ``{Constraining Dark
  Matter-Neutrino Interactions using the CMB and Large-Scale Structure},''
  \href{http://dx.doi.org/10.1088/1475-7516/2014/05/011}{{\em JCAP} {\bfseries
  05} (2014) 011}, \href{http://arxiv.org/abs/1401.7597}{{\ttfamily
  arXiv:1401.7597 [astro-ph.CO]}}.

\bibitem{Zhang:2020nis}
Y.~Zhang, ``{Speeding up dark matter with solar neutrinos},''
  \href{http://dx.doi.org/10.1093/ptep/ptab156}{{\em PTEP} {\bfseries 2022}
  no.~1, (2022) 013B05}, \href{http://arxiv.org/abs/2001.00948}{{\ttfamily
  arXiv:2001.00948 [hep-ph]}}.

\bibitem{Farzan:2014gza}
Y.~Farzan and S.~Palomares-Ruiz, ``{Dips in the Diffuse Supernova Neutrino
  Background},'' \href{http://dx.doi.org/10.1088/1475-7516/2014/06/014}{{\em
  JCAP} {\bfseries 06} (2014) 014},
  \href{http://arxiv.org/abs/1401.7019}{{\ttfamily arXiv:1401.7019 [hep-ph]}}.

\bibitem{Das:2021lcr}
A.~Das and M.~Sen, ``{Boosted dark matter from diffuse supernova neutrinos},''
  \href{http://dx.doi.org/10.1103/PhysRevD.104.075029}{{\em Phys. Rev. D}
  {\bfseries 104} no.~7, (2021) 075029},
  \href{http://arxiv.org/abs/2104.00027}{{\ttfamily arXiv:2104.00027
  [hep-ph]}}.

\bibitem{Bardhan:2022ywd}
D.~Bardhan, S.~Bhowmick, D.~Ghosh, A.~Guha, and D.~Sachdeva, ``{Boosting
  through the Darkness},'' \href{http://arxiv.org/abs/2208.09405}{{\ttfamily
  arXiv:2208.09405 [hep-ph]}}.

\bibitem{Fayet:2006sa}
P.~Fayet, D.~Hooper, and G.~Sigl, ``{Constraints on light dark matter from
  core-collapse supernovae},''
  \href{http://dx.doi.org/10.1103/PhysRevLett.96.211302}{{\em Phys. Rev. Lett.}
  {\bfseries 96} (2006) 211302},
  \href{http://arxiv.org/abs/hep-ph/0602169}{{\ttfamily arXiv:hep-ph/0602169}}.

\bibitem{McMullen:2021ikf}
{\bfseries IceCube} Collaboration, A.~McMullen, A.~Vincent, C.~Arguelles, and
  A.~Schneider, ``{Dark matter neutrino scattering in the galactic centre with
  IceCube},'' \href{http://dx.doi.org/10.1088/1748-0221/16/08/C08001}{{\em
  JINST} {\bfseries 16} no.~08, (2021) C08001},
  \href{http://arxiv.org/abs/2107.11491}{{\ttfamily arXiv:2107.11491
  [astro-ph.HE]}}.

\bibitem{Koren:2019wwi}
S.~Koren, ``{Neutrino -- Dark Matter Scattering and Coincident Detections of
  UHE Neutrinos with EM Sources},''
  \href{http://dx.doi.org/10.1088/1475-7516/2019/09/013}{{\em JCAP} {\bfseries
  09} (2019) 013}, \href{http://arxiv.org/abs/1903.05096}{{\ttfamily
  arXiv:1903.05096 [hep-ph]}}.

\bibitem{Murase:2019xqi}
K.~Murase and I.~M. Shoemaker, ``{Neutrino Echoes from Multimessenger Transient
  Sources},'' \href{http://dx.doi.org/10.1103/PhysRevLett.123.241102}{{\em
  Phys. Rev. Lett.} {\bfseries 123} no.~24, (2019) 241102},
  \href{http://arxiv.org/abs/1903.08607}{{\ttfamily arXiv:1903.08607
  [hep-ph]}}.

\bibitem{Carpio:2022sml}
J.~A. Carpio, A.~Kheirandish, and K.~Murase, ``{Time-delayed neutrino emission
  from supernovae as a probe of dark matter-neutrino interactions},''
  \href{http://arxiv.org/abs/2204.09650}{{\ttfamily arXiv:2204.09650
  [hep-ph]}}.

\bibitem{Davis:2015rza}
J.~H. Davis and J.~Silk, ``{Spectral and Spatial Distortions of PeV Neutrinos
  from Scattering with Dark Matter},''
  \href{http://arxiv.org/abs/1505.01843}{{\ttfamily arXiv:1505.01843
  [hep-ph]}}.

\bibitem{Yin:2018yjn}
W.~Yin, ``{Highly-boosted dark matter and cutoff for cosmic-ray neutrinos
  through neutrino portal},''
  \href{http://dx.doi.org/10.1051/epjconf/201920804003}{{\em EPJ Web Conf.}
  {\bfseries 208} (2019) 04003},
  \href{http://arxiv.org/abs/1809.08610}{{\ttfamily arXiv:1809.08610
  [hep-ph]}}.

\bibitem{Dvorkin:2013cea}
C.~Dvorkin, K.~Blum, and M.~Kamionkowski, ``{Constraining Dark Matter-Baryon
  Scattering with Linear Cosmology},''
  \href{http://dx.doi.org/10.1103/PhysRevD.89.023519}{{\em Phys. Rev. D}
  {\bfseries 89} no.~2, (2014) 023519},
  \href{http://arxiv.org/abs/1311.2937}{{\ttfamily arXiv:1311.2937
  [astro-ph.CO]}}.

\bibitem{Buen-Abad:2021mvc}
M.~A. Buen-Abad, R.~Essig, D.~McKeen, and Y.-M. Zhong, ``{Cosmological
  constraints on dark matter interactions with ordinary matter},''
  \href{http://dx.doi.org/10.1016/j.physrep.2022.02.006}{{\em Phys. Rept.}
  {\bfseries 961} (2022) 1--35},
  \href{http://arxiv.org/abs/2107.12377}{{\ttfamily arXiv:2107.12377
  [astro-ph.CO]}}.

\bibitem{Nguyen:2021cnb}
D.~V. Nguyen, D.~Sarnaaik, K.~K. Boddy, E.~O. Nadler, and V.~Gluscevic,
  ``{Observational constraints on dark matter scattering with electrons},''
  \href{http://dx.doi.org/10.1103/PhysRevD.104.103521}{{\em Phys. Rev. D}
  {\bfseries 104} no.~10, (2021) 103521},
  \href{http://arxiv.org/abs/2107.12380}{{\ttfamily arXiv:2107.12380
  [astro-ph.CO]}}.

\bibitem{Ali-Haimoud:2015pwa}
Y.~Ali-Ha\"\i{}moud, J.~Chluba, and M.~Kamionkowski, ``{Constraints on Dark
  Matter Interactions with Standard Model Particles from Cosmic Microwave
  Background Spectral Distortions},''
  \href{http://dx.doi.org/10.1103/PhysRevLett.115.071304}{{\em Phys. Rev.
  Lett.} {\bfseries 115} no.~7, (2015) 071304},
  \href{http://arxiv.org/abs/1506.04745}{{\ttfamily arXiv:1506.04745
  [astro-ph.CO]}}.

\bibitem{Ali-Haimoud:2021lka}
Y.~Ali-Ha\"\i{}moud, ``{Testing dark matter interactions with CMB spectral
  distortions},'' \href{http://dx.doi.org/10.1103/PhysRevD.103.043541}{{\em
  Phys. Rev. D} {\bfseries 103} no.~4, (2021) 043541},
  \href{http://arxiv.org/abs/2101.04070}{{\ttfamily arXiv:2101.04070
  [astro-ph.CO]}}.

\bibitem{Wadekar:2019mpc}
D.~Wadekar and G.~R. Farrar, ``{Gas-rich dwarf galaxies as a new probe of dark
  matter interactions with ordinary matter},''
  \href{http://dx.doi.org/10.1103/PhysRevD.103.123028}{{\em Phys. Rev. D}
  {\bfseries 103} no.~12, (2021) 123028},
  \href{http://arxiv.org/abs/1903.12190}{{\ttfamily arXiv:1903.12190
  [hep-ph]}}.

\bibitem{Cappiello:2019qsw}
C.~V. Cappiello and J.~F. Beacom, ``{Strong New Limits on Light Dark Matter
  from Neutrino Experiments},''
  \href{http://dx.doi.org/10.1103/PhysRevD.104.069901}{{\em Phys. Rev. D}
  {\bfseries 100} no.~10, (2019) 103011},
  \href{http://arxiv.org/abs/1906.11283}{{\ttfamily arXiv:1906.11283
  [hep-ph]}}. [Erratum: Phys.Rev.D 104, 069901 (2021)].

\bibitem{Ema:2018bih}
Y.~Ema, F.~Sala, and R.~Sato, ``{Light Dark Matter at Neutrino Experiments},''
  \href{http://dx.doi.org/10.1103/PhysRevLett.122.181802}{{\em Phys. Rev.
  Lett.} {\bfseries 122} no.~18, (2019) 181802},
  \href{http://arxiv.org/abs/1811.00520}{{\ttfamily arXiv:1811.00520
  [hep-ph]}}.

\bibitem{Cao:2020bwd}
Q.-H. Cao, R.~Ding, and Q.-F. Xiang, ``{Searching for sub-MeV boosted dark
  matter from xenon electron direct detection},''
  \href{http://dx.doi.org/10.1088/1674-1137/abe195}{{\em Chin. Phys. C}
  {\bfseries 45} no.~4, (2021) 045002},
  \href{http://arxiv.org/abs/2006.12767}{{\ttfamily arXiv:2006.12767
  [hep-ph]}}.

\bibitem{Xia:2020apm}
C.~Xia, Y.-H. Xu, and Y.-F. Zhou, ``{Constraining light dark matter upscattered
  by ultrahigh-energy cosmic rays},''
  \href{http://dx.doi.org/10.1016/j.nuclphysb.2021.115470}{{\em Nucl. Phys. B}
  {\bfseries 969} (2021) 115470},
  \href{http://arxiv.org/abs/2009.00353}{{\ttfamily arXiv:2009.00353
  [hep-ph]}}.

\bibitem{Dent:2020syp}
J.~B. Dent, B.~Dutta, J.~L. Newstead, I.~M. Shoemaker, and N.~T. Arellano,
  ``{Present and future status of light dark matter models from cosmic-ray
  electron upscattering},''
  \href{http://dx.doi.org/10.1103/PhysRevD.103.095015}{{\em Phys. Rev. D}
  {\bfseries 103} (2021) 095015},
  \href{http://arxiv.org/abs/2010.09749}{{\ttfamily arXiv:2010.09749
  [hep-ph]}}.

\bibitem{Bringmann:2018cvk}
T.~Bringmann and M.~Pospelov, ``{Novel direct detection constraints on light
  dark matter},'' \href{http://dx.doi.org/10.1103/PhysRevLett.122.171801}{{\em
  Phys. Rev. Lett.} {\bfseries 122} no.~17, (2019) 171801},
  \href{http://arxiv.org/abs/1810.10543}{{\ttfamily arXiv:1810.10543
  [hep-ph]}}.

\bibitem{Dent:2019krz}
J.~B. Dent, B.~Dutta, J.~L. Newstead, and I.~M. Shoemaker, ``{Bounds on Cosmic
  Ray-Boosted Dark Matter in Simplified Models and its Corresponding
  Neutrino-Floor},'' \href{http://dx.doi.org/10.1103/PhysRevD.101.116007}{{\em
  Phys. Rev. D} {\bfseries 101} no.~11, (2020) 116007},
  \href{http://arxiv.org/abs/1907.03782}{{\ttfamily arXiv:1907.03782
  [hep-ph]}}.

\bibitem{Wang:2019jtk}
W.~Wang, L.~Wu, J.~M. Yang, H.~Zhou, and B.~Zhu, ``{Cosmic ray boosted sub-GeV
  gravitationally interacting dark matter in direct detection},''
  \href{http://dx.doi.org/10.1007/JHEP12(2020)072}{{\em JHEP} {\bfseries 12}
  (2020) 072}, \href{http://arxiv.org/abs/1912.09904}{{\ttfamily
  arXiv:1912.09904 [hep-ph]}}. [Erratum: JHEP 02, 052 (2021)].

\bibitem{Guo:2020drq}
G.~Guo, Y.-L.~S. Tsai, and M.-R. Wu, ``{Probing cosmic-ray accelerated light
  dark matter with IceCube},''
  \href{http://dx.doi.org/10.1088/1475-7516/2020/10/049}{{\em JCAP} {\bfseries
  10} (2020) 049}, \href{http://arxiv.org/abs/2004.03161}{{\ttfamily
  arXiv:2004.03161 [astro-ph.HE]}}.

\bibitem{Ge:2020yuf}
S.-F. Ge, J.~Liu, Q.~Yuan, and N.~Zhou, ``{Diurnal Effect of Sub-GeV Dark
  Matter Boosted by Cosmic Rays},''
  \href{http://dx.doi.org/10.1103/PhysRevLett.126.091804}{{\em Phys. Rev.
  Lett.} {\bfseries 126} no.~9, (2021) 091804},
  \href{http://arxiv.org/abs/2005.09480}{{\ttfamily arXiv:2005.09480
  [hep-ph]}}.

\bibitem{Jho:2020sku}
Y.~Jho, J.-C. Park, S.~C. Park, and P.-Y. Tseng, ``{Leptonic New Force and
  Cosmic-ray Boosted Dark Matter for the XENON1T Excess},''
  \href{http://dx.doi.org/10.1016/j.physletb.2020.135863}{{\em Phys. Lett. B}
  {\bfseries 811} (2020) 135863},
  \href{http://arxiv.org/abs/2006.13910}{{\ttfamily arXiv:2006.13910
  [hep-ph]}}.

\bibitem{Cho:2020mnc}
W.~Cho, K.-Y. Choi, and S.~M. Yoo, ``{Searching for boosted dark matter
  mediated by a new gauge boson},''
  \href{http://dx.doi.org/10.1103/PhysRevD.102.095010}{{\em Phys. Rev. D}
  {\bfseries 102} no.~9, (2020) 095010},
  \href{http://arxiv.org/abs/2007.04555}{{\ttfamily arXiv:2007.04555
  [hep-ph]}}.

\bibitem{Lei:2020mii}
Z.-H. Lei, J.~Tang, and B.-L. Zhang, ``{Constraints on cosmic-ray boosted dark
  matter in CDEX-10 *},''
  \href{http://dx.doi.org/10.1088/1674-1137/ac68da}{{\em Chin. Phys. C}
  {\bfseries 46} no.~8, (2022) 085103},
  \href{http://arxiv.org/abs/2008.07116}{{\ttfamily arXiv:2008.07116
  [hep-ph]}}.

\bibitem{Guo:2020oum}
G.~Guo, Y.-L.~S. Tsai, M.-R. Wu, and Q.~Yuan, ``{Elastic and Inelastic
  Scattering of Cosmic-Rays on Sub-GeV Dark Matter},''
  \href{http://dx.doi.org/10.1103/PhysRevD.102.103004}{{\em Phys. Rev. D}
  {\bfseries 102} no.~10, (2020) 103004},
  \href{http://arxiv.org/abs/2008.12137}{{\ttfamily arXiv:2008.12137
  [astro-ph.HE]}}.

\bibitem{Ema:2020ulo}
Y.~Ema, F.~Sala, and R.~Sato, ``{Neutrino experiments probe hadrophilic light
  dark matter},'' \href{http://dx.doi.org/10.21468/SciPostPhys.10.3.072}{{\em
  SciPost Phys.} {\bfseries 10} no.~3, (2021) 072},
  \href{http://arxiv.org/abs/2011.01939}{{\ttfamily arXiv:2011.01939
  [hep-ph]}}.

\bibitem{Flambaum:2020xxo}
V.~V. Flambaum, L.~Su, L.~Wu, and B.~Zhu, ``{Constraining sub-GeV dark matter
  from Migdal and Boosted effects},''
  \href{http://arxiv.org/abs/2012.09751}{{\ttfamily arXiv:2012.09751
  [hep-ph]}}.

\bibitem{Bell:2021xff}
N.~F. Bell, J.~B. Dent, B.~Dutta, S.~Ghosh, J.~Kumar, J.~L. Newstead, and I.~M.
  Shoemaker, ``{Cosmic-ray upscattered inelastic dark matter},''
  \href{http://dx.doi.org/10.1103/PhysRevD.104.076020}{{\em Phys. Rev. D}
  {\bfseries 104} (2021) 076020},
  \href{http://arxiv.org/abs/2108.00583}{{\ttfamily arXiv:2108.00583
  [hep-ph]}}.

\bibitem{Feng:2021hyz}
J.-C. Feng, X.-W. Kang, C.-T. Lu, Y.-L.~S. Tsai, and F.-S. Zhang, ``{Revising
  inelastic dark matter direct detection by including the cosmic ray
  acceleration},'' \href{http://dx.doi.org/10.1007/JHEP04(2022)080}{{\em JHEP}
  {\bfseries 04} (2022) 080}, \href{http://arxiv.org/abs/2110.08863}{{\ttfamily
  arXiv:2110.08863 [hep-ph]}}.

\bibitem{Wang:2021nbf}
W.~Wang, L.~Wu, W.-N. Yang, and B.~Zhu, ``{The Spin-dependent Scattering of
  Boosted Dark Matter},'' \href{http://arxiv.org/abs/2111.04000}{{\ttfamily
  arXiv:2111.04000 [hep-ph]}}.

\bibitem{Xia:2021vbz}
C.~Xia, Y.-H. Xu, and Y.-F. Zhou, ``{Production and attenuation of cosmic-ray
  boosted dark matter},''
  \href{http://dx.doi.org/10.1088/1475-7516/2022/02/028}{{\em JCAP} {\bfseries
  02} no.~02, (2022) 028}, \href{http://arxiv.org/abs/2111.05559}{{\ttfamily
  arXiv:2111.05559 [hep-ph]}}.

\bibitem{Bramante:2021dyx}
J.~Bramante, B.~J. Kavanagh, and N.~Raj, ``{Scattering Searches for Dark Matter
  in Subhalos: Neutron Stars, Cosmic Rays, and Old Rocks},''
  \href{http://dx.doi.org/10.1103/PhysRevLett.128.231801}{{\em Phys. Rev.
  Lett.} {\bfseries 128} no.~23, (2022) 231801},
  \href{http://arxiv.org/abs/2109.04582}{{\ttfamily arXiv:2109.04582
  [hep-ph]}}.

\bibitem{Essig:2012yx}
R.~Essig, A.~Manalaysay, J.~Mardon, P.~Sorensen, and T.~Volansky, ``{First
  Direct Detection Limits on sub-GeV Dark Matter from XENON10},''
  \href{http://dx.doi.org/10.1103/PhysRevLett.109.021301}{{\em Phys. Rev.
  Lett.} {\bfseries 109} (2012) 021301},
  \href{http://arxiv.org/abs/1206.2644}{{\ttfamily arXiv:1206.2644
  [astro-ph.CO]}}.

\bibitem{SuperCDMS:2018mne}
{\bfseries SuperCDMS} Collaboration, R.~Agnese {\em et~al.}, ``{First Dark
  Matter Constraints from a SuperCDMS Single-Charge Sensitive Detector},''
  \href{http://dx.doi.org/10.1103/PhysRevLett.121.051301}{{\em Phys. Rev.
  Lett.} {\bfseries 121} no.~5, (2018) 051301},
  \href{http://arxiv.org/abs/1804.10697}{{\ttfamily arXiv:1804.10697
  [hep-ex]}}. [Erratum: Phys.Rev.Lett. 122, 069901 (2019)].

\bibitem{SuperCDMS:2020ymb}
{\bfseries SuperCDMS} Collaboration, D.~W. Amaral {\em et~al.}, ``{Constraints
  on low-mass, relic dark matter candidates from a surface-operated SuperCDMS
  single-charge sensitive detector},''
  \href{http://dx.doi.org/10.1103/PhysRevD.102.091101}{{\em Phys. Rev. D}
  {\bfseries 102} no.~9, (2020) 091101},
  \href{http://arxiv.org/abs/2005.14067}{{\ttfamily arXiv:2005.14067
  [hep-ex]}}.

\bibitem{DarkSide:2018ppu}
{\bfseries DarkSide} Collaboration, P.~Agnes {\em et~al.}, ``{Constraints on
  Sub-GeV Dark-Matter\textendash{}Electron Scattering from the DarkSide-50
  Experiment},'' \href{http://dx.doi.org/10.1103/PhysRevLett.121.111303}{{\em
  Phys. Rev. Lett.} {\bfseries 121} no.~11, (2018) 111303},
  \href{http://arxiv.org/abs/1802.06998}{{\ttfamily arXiv:1802.06998
  [astro-ph.CO]}}.

\bibitem{Crisler:2018gci}
{\bfseries SENSEI} Collaboration, M.~Crisler, R.~Essig, J.~Estrada,
  G.~Fernandez, J.~Tiffenberg, M.~Sofo~haro, T.~Volansky, and T.-T. Yu,
  ``{SENSEI: First Direct-Detection Constraints on sub-GeV Dark Matter from a
  Surface Run},'' \href{http://dx.doi.org/10.1103/PhysRevLett.121.061803}{{\em
  Phys. Rev. Lett.} {\bfseries 121} no.~6, (2018) 061803},
  \href{http://arxiv.org/abs/1804.00088}{{\ttfamily arXiv:1804.00088
  [hep-ex]}}.

\bibitem{SENSEI:2019ibb}
{\bfseries SENSEI} Collaboration, O.~Abramoff {\em et~al.}, ``{SENSEI:
  Direct-Detection Constraints on Sub-GeV Dark Matter from a Shallow
  Underground Run Using a Prototype Skipper-CCD},''
  \href{http://dx.doi.org/10.1103/PhysRevLett.122.161801}{{\em Phys. Rev.
  Lett.} {\bfseries 122} no.~16, (2019) 161801},
  \href{http://arxiv.org/abs/1901.10478}{{\ttfamily arXiv:1901.10478
  [hep-ex]}}.

\bibitem{XENON:2021nad}
{\bfseries XENON} Collaboration, E.~Aprile {\em et~al.}, ``{Emission of single
  and few electrons in XENON1T and limits on light dark matter},''
  \href{http://dx.doi.org/10.1103/PhysRevD.106.022001}{{\em Phys. Rev. D}
  {\bfseries 106} no.~2, (2022) 022001},
  \href{http://arxiv.org/abs/2112.12116}{{\ttfamily arXiv:2112.12116
  [hep-ex]}}.

\bibitem{DAMIC:2019dcn}
{\bfseries DAMIC} Collaboration, A.~Aguilar-Arevalo {\em et~al.},
  ``{Constraints on Light Dark Matter Particles Interacting with Electrons from
  DAMIC at SNOLAB},''
  \href{http://dx.doi.org/10.1103/PhysRevLett.123.181802}{{\em Phys. Rev.
  Lett.} {\bfseries 123} no.~18, (2019) 181802},
  \href{http://arxiv.org/abs/1907.12628}{{\ttfamily arXiv:1907.12628
  [astro-ph.CO]}}.

\bibitem{EDELWEISS:2020fxc}
{\bfseries EDELWEISS} Collaboration, Q.~Arnaud {\em et~al.}, ``{First
  germanium-based constraints on sub-MeV Dark Matter with the EDELWEISS
  experiment},'' \href{http://dx.doi.org/10.1103/PhysRevLett.125.141301}{{\em
  Phys. Rev. Lett.} {\bfseries 125} no.~14, (2020) 141301},
  \href{http://arxiv.org/abs/2003.01046}{{\ttfamily arXiv:2003.01046
  [astro-ph.GA]}}.

\bibitem{PandaX-II:2021nsg}
{\bfseries PandaX-II} Collaboration, C.~Cheng {\em et~al.}, ``{Search for Light
  Dark Matter-Electron Scatterings in the PandaX-II Experiment},''
  \href{http://dx.doi.org/10.1103/PhysRevLett.126.211803}{{\em Phys. Rev.
  Lett.} {\bfseries 126} no.~21, (2021) 211803},
  \href{http://arxiv.org/abs/2101.07479}{{\ttfamily arXiv:2101.07479
  [hep-ex]}}.

\bibitem{Cappiello:2018hsu}
C.~V. Cappiello, K.~C.~Y. Ng, and J.~F. Beacom, ``{Reverse Direct Detection:
  Cosmic Ray Scattering With Light Dark Matter},''
  \href{http://dx.doi.org/10.1103/PhysRevD.99.063004}{{\em Phys. Rev. D}
  {\bfseries 99} no.~6, (2019) 063004},
  \href{http://arxiv.org/abs/1810.07705}{{\ttfamily arXiv:1810.07705
  [hep-ph]}}.

\bibitem{Bell:2019pyc}
N.~F. Bell, G.~Busoni, and S.~Robles, ``{Capture of Leptophilic Dark Matter in
  Neutron Stars},'' \href{http://dx.doi.org/10.1088/1475-7516/2019/06/054}{{\em
  JCAP} {\bfseries 06} (2019) 054},
  \href{http://arxiv.org/abs/1904.09803}{{\ttfamily arXiv:1904.09803
  [hep-ph]}}.

\bibitem{Bell:2020jou}
N.~F. Bell, G.~Busoni, S.~Robles, and M.~Virgato, ``{Improved Treatment of Dark
  Matter Capture in Neutron Stars},''
  \href{http://dx.doi.org/10.1088/1475-7516/2020/09/028}{{\em JCAP} {\bfseries
  09} (2020) 028}, \href{http://arxiv.org/abs/2004.14888}{{\ttfamily
  arXiv:2004.14888 [hep-ph]}}.

\bibitem{Bell:2020lmm}
N.~F. Bell, G.~Busoni, S.~Robles, and M.~Virgato, ``{Improved Treatment of Dark
  Matter Capture in Neutron Stars II: Leptonic Targets},''
  \href{http://dx.doi.org/10.1088/1475-7516/2021/03/086}{{\em JCAP} {\bfseries
  03} (2021) 086}, \href{http://arxiv.org/abs/2010.13257}{{\ttfamily
  arXiv:2010.13257 [hep-ph]}}.

\bibitem{Bell:2021fye}
N.~F. Bell, G.~Busoni, M.~E. Ramirez-Quezada, S.~Robles, and M.~Virgato,
  ``{Improved treatment of dark matter capture in white dwarfs},''
  \href{http://dx.doi.org/10.1088/1475-7516/2021/10/083}{{\em JCAP} {\bfseries
  10} (2021) 083}, \href{http://arxiv.org/abs/2104.14367}{{\ttfamily
  arXiv:2104.14367 [hep-ph]}}.

\bibitem{Bilenky:1992xn}
M.~S. Bilenky, S.~M. Bilenky, and A.~Santamaria, ``{Invisible width of the Z
  boson and 'secret' neutrino-neutrino interactions},''
  \href{http://dx.doi.org/10.1016/0370-2693(93)90703-K}{{\em Phys. Lett. B}
  {\bfseries 301} (1993) 287--291}.

\bibitem{Berryman:2022hds}
J.~M. Berryman {\em et~al.}, ``{Neutrino Self-Interactions: A White Paper},''
  in {\em {2022 Snowmass Summer Study}}.
\newblock 3, 2022.
\newblock \href{http://arxiv.org/abs/2203.01955}{{\ttfamily arXiv:2203.01955
  [hep-ph]}}.

\bibitem{inprogress}
J.~M. Cline and M.~Puel, ``{NGC 1068 constraints on neutrino-dark matter
  scattering},'' {\em in preparation} (2023) .

\end{thebibliography}\endgroup
\end{document}